\documentclass[a4paper,aps,prd,10pt,preprintnumbers,showpacs,twocolumn,superscriptaddress,nofootinbib,amsmath,amssymb]{revtex4-1}
\usepackage{graphicx}
\usepackage{cmap}
\usepackage[utf8]{inputenc}
\usepackage[T1]{fontenc}

\def\imo{i}

\def\K{{\cal K}}

\begin{document}
\title{Quasinormal Frequencies of Fields with Various Spin in the Quantum Oppenheimer-Snyder Model of Black Holes}
\author{Milena Skvortsova}\email{milenas577@mail.ru}
\affiliation{Peoples' Friendship University of Russia (RUDN University), 6 Miklukho-Maklaya Street, Moscow, 117198, Russia}
\begin{abstract}
A recent development involves an intriguing model of a quantum-corrected black hole, established through the application of the quantum Oppenheimer-Snyder model within loop quantum cosmology [Lewandowski et al., Phys. Rev. Lett. 130 (2023) 10, 101501]. Employing both time-domain integration and the WKB approach, we compute the quasinormal frequencies for scalar, electromagnetic, and neutrino perturbations in these quantum-corrected black holes. Our analysis reveals that while the real oscillation frequencies undergo only minor adjustments due to the quantum parameter, the damping rate experiences a significant decrease as a result of its influence. We also deduce the analytic formula for quasinormal frequencies in the eikonal limit and show that the correspondence between the null geodesics and eikoanl quasinormal modes holds in this case.
\end{abstract}
\maketitle
\section{Introduction}

Quasinormal modes of black holes represent the characteristic oscillations of spacetime occurring after a perturbation, such as the merger of two black holes or the infall of matter. These modes are crucial in gravitational wave astronomy as they encode information about the black hole's properties, including its mass, spin, and geometry.
Quasinormal frequencies have been detected using gravitational interferometers of the LIGO/VIRGO collaboration \cite{LIGOScientific:2016aoc,LIGOScientific:2017vwq,LIGOScientific:2020zkf}, and upcoming experiments hold the promise of a significantly broader frequency range for observations \cite{Babak:2017tow}. Concurrently, ongoing and future observations of black holes across the electromagnetic spectrum \cite{EventHorizonTelescope:2019dse,Goddi:2016qax} may provide more robust constraints on black hole geometry \cite{Tsukamoto:2014tja,Shaikh:2021yux}. However, the uncertainty surrounding the mass and angular momentum of the final black hole allows for the possibility of attributing the observed signal to a wide range of non-Kerr spacetimes, thus leaving space for various alternative theories of gravity.

An extensive area of search for alternatives to the Einstein gravity is related to introduction, in various ways, of quantum corrections to the black hole spacetime.  
Perturbations, scattering properties and quasinormal spectrum of black hole geometries with various types of quantum corrections have been extensively analyzed in \cite{Konoplya:2019xmn,Bronnikov:2023uuv,Konoplya:2023ahd,Karmakar:2022idu,Bolokhov:2023bwm,Cruz:2020emz,Konoplya:2022hll,Yang:2022btw,Liu:2020ola,Chen:2022ngd,Konoplya:2023ahd,Daghigh:2020fmw,Baruah:2023rhd,Saleh:2016pke}.
Nevertheless, there are gaps in the literature on quasinormal spectrum of the quantum corrected black holes.

An interesting model in this respect has been considered in the framework of the quantum Oppenheimer-Snyder model in loop quantum cosmology \cite{Lewandowski:2022zce}.  There, a new
quantum black hole model was suggested whose metric tensor is a suitably deformed Schwarzschild's metric.
The quantum effects considered in \cite{Lewandowski:2022zce} lead to a lower bound on the mass of the black hole produced during the collapse of a
dust ball. The quasinormal modes of a test scalar field has been recently considered in  \cite{Gong:2023ghh} with the WKB method. Here, we complement this work by consideration also electromagnetic and neutrino perturbations, and also show that some of the results presented in  \cite{Gong:2023ghh} suffer from insufficient accuracy. In addition to massless fields, we compute quasinormal modes of a massive scalar field and show that there must be arbitrarily long lived quasinormal frequencies in the spectrum, because the damping rates tend to the Schwarzschild ones, when the field's mass $\mu$ is increased.

The paper is organized as follows. In sec.  \ref{sec:wavelike} we describe the metric, wave-like euqations and effective potentials. Sec. \ref{sec:methods} reviews the methods used for calculations of quasinormal modes: time-domain integration method with the Prony procedure for extracting frequencies and the higher order WKB approach with Padé approximants. Sec. \ref{sec:QNM} is devoted to calculations of quasinormal modes for scalar, electromagnetic and Dirac fields. Finally, in sec.  \ref{sec:conclusions} we briefly summarize the obtained results.

\section{Basic equations}\label{sec:wavelike}

The metric of the quantum corrected black hole has the following form \cite{Lewandowski:2022zce},
\begin{equation}\label{metric}
  ds^2=-f(r)dt^2+\frac{dr^2}{f(r)}+r^2(d\theta^2+\sin^2\theta d\phi^2),
\end{equation}
where
$$
\begin{array}{rcl}
f(r)&=&\displaystyle 1-\frac{2 M}{r} + \frac{\gamma  M^2}{r^4}.\\
\end{array}
$$
Here $\gamma $ is the quantum parameter, connected to the Immirzi parameter and Plank length, $M$ is the ADM mass. We will measure all dimensional quantities in units of the mass $M=1$. The event horizon exists once $0 \leq  \alpha \leq 26/17$.

The general covariant equations for test scalar ($\Phi$), electromagnetic ($A_\mu$), and Dirac ($\Upsilon$) fields have the form:
\begin{subequations}\label{coveqs}
\begin{eqnarray}\label{KGg}
\frac{1}{\sqrt{-g}}\partial_\mu \left(\sqrt{-g}g^{\mu \nu}\partial_\nu\Phi\right)&=&\mu^2 \Psi,
\\\label{EmagEq}
\frac{1}{\sqrt{-g}}\partial_{\mu} \left(F_{\rho\sigma}g^{\rho \nu}g^{\sigma \mu}\sqrt{-g}\right)&=&0\,,
\\\label{covdirac}
\gamma^{\alpha} \left( \frac{\partial}{\partial x^{\alpha}} - \Gamma_{\alpha} \right) \Upsilon&=&0,
\end{eqnarray}
\end{subequations}
where $F_{\mu\nu}=\partial_\mu A_\nu-\partial_\nu A_\mu$ is the electromagnetic tensor, $\gamma^{\alpha}$ are noncommutative gamma matrices and $\Gamma_{\alpha}$ are tetrad spin connections.
After separation of the variables, the dynamical equations for test fields (\ref{coveqs}) can be reduced to  the following wavelike form \cite{Kokkotas:1999bd,Berti:2009kk,Konoplya:2011qq}:
\begin{equation}\label{wave-equation}
\dfrac{d^2 \Psi}{dr_*^2}+(\omega^2-V(r))\Psi=0,
\end{equation}
where the ``tortoise coordinate'' $r_*$ is
\begin{equation}\label{tortoise}
dr_*\equiv\frac{dr}{f(r)}.
\end{equation}

The effective potentials for scalar ($s=0$) and electromagnetic ($s=1$) fields have the following form
\begin{equation}\label{potentialScalar}
V(r)=f(r)\frac{\ell(\ell+1)}{r^2}+\frac{1-s}{r}\cdot\frac{d^2 r}{dr_*^2}.
\end{equation}
In the case of massive scalar field the term $f(r) \mu^2$ is added to the effective potential of the massless scalar field.
Here $\ell=s, s+1, s+2, \ldots$ are multipole numbers, coming from the separation of the angular variables.
For the Dirac field ($s=1/2$) we obtain the two isospectral potentials,
\begin{equation}
V_{\pm}(r) = W^2\pm\frac{dW}{dr_*}, \quad W\equiv \left(\ell+\frac{1}{2}\right)\frac{\sqrt{f(r)}}{r}.
\end{equation}
The isospectral wave functions can be transformed one into another by the Darboux transformation,
\begin{equation}\label{psi}
\Psi_{+}\propto \left(W+\dfrac{d}{dr_*}\right) \Psi_{-}.
\end{equation}
Thus, we will study only one of the effective potentials, $V_{+}(r)$, because the WKB method is usually more accurate in this case.

\begin{figure}
\resizebox{\linewidth}{!}{\includegraphics{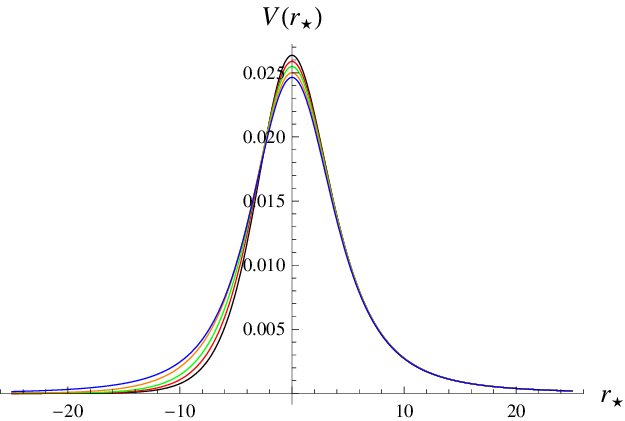}}
\caption{Potential as a function of the tortoise coordinate $r_{*}$ for the $\ell=0$ scalar field  ($M=1$): $\gamma=0$ (black), $\gamma=0.6$ (red), $\gamma=1$ (green), $\gamma=1.4$ (orange), $\gamma=1.67$ (blue).}\label{fig:potentials}
\end{figure}

\begin{figure}
\resizebox{\linewidth}{!}{\includegraphics{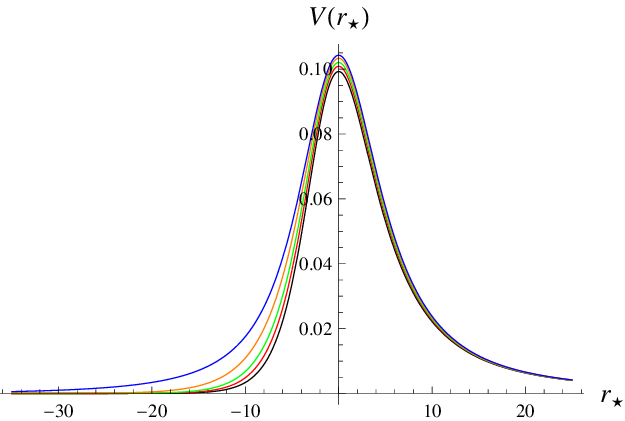}}
\caption{Potential as a function of the tortoise coordinate $r_{*}$ for the $\ell=1$ scalar field  ($M=1$): $\gamma=0$ (black), $\gamma=0.6$ (red), $\gamma=1$ (green), $\gamma=1.4$ (orange), $\gamma=1.67$ (blue).}\label{fig:potentials2}
\end{figure}

\begin{figure}
\resizebox{\linewidth}{!}{\includegraphics{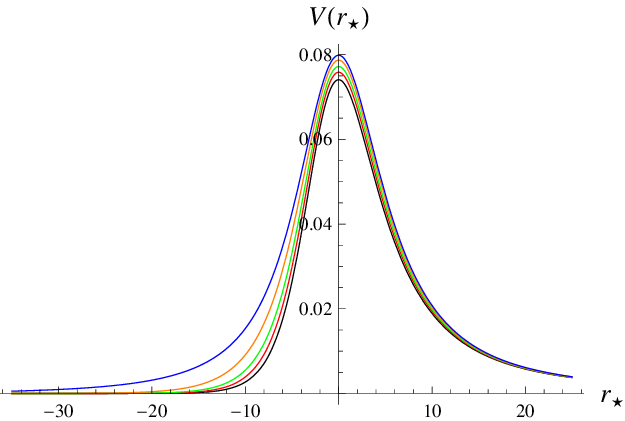}}
\caption{Potential as a function of the tortoise coordinate $r_{*}$ for the $\ell=1$ scalar field  ($M=1$): $\gamma=0$ (black), $\gamma=0.6$ (red), $\gamma=1$ (green), $\gamma=1.4$ (orange), $\gamma=1.67$ (blue).}\label{fig:potentials3}
\end{figure}

\begin{figure}
\resizebox{\linewidth}{!}{\includegraphics{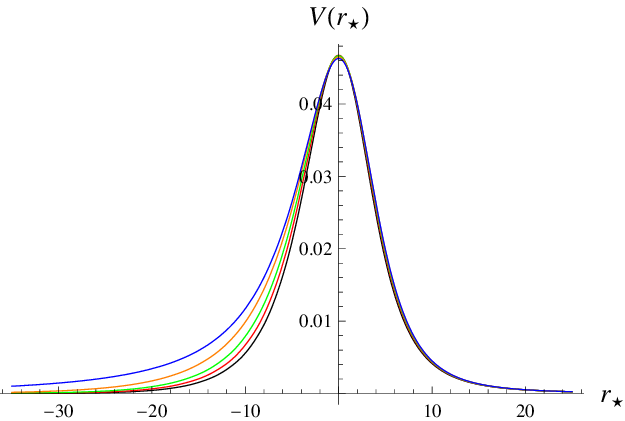}}
\caption{Potential as a function of the tortoise coordinate $r_{*}$ for the $\ell=1/2$ scalar field  ($M=1$, $\Lambda=1/10$): $\gamma=0$ (black), $\gamma=0.6$ (red), $\gamma=1$ (green), $\gamma=1.4$ (orange), $\gamma=1.67$ (blue).}\label{fig:potentials4}
\end{figure}

Examples of the effective potentials are given in figs. \ref{fig:potentials}-\ref{fig:potentials4}. The effective potential are positive definite, which guarantees the stability of the field against small perturbations. The larger values of the quantum correction parameter $\gamma$ corresponds to the higher effective potentials.

\section{WKB method, time-domain integration and Prony method}\label{sec:methods}

Qausinormal modes $\omega$ are complex values of the wave-like equation \ref{wave-equation} under specific boundary conditions: purely ingoing waves at the event horizon ($r_*\to-\infty$) and purely outgoing waves at infinity ($r_*\to\infty$),
\begin{equation}\label{boundaryconditions}
\Psi(r_*\to\pm\infty)\propto e^{\pm\imo \omega r_*}.
\end{equation}
Here we will consider two independent methods for calculations quasinormal modes: WKB approach and the time-domain integration. 

\subsection{WKB method}

When the effective potential $V(r)$ in the wave equation (\ref{wave-equation}) takes the form of a barrier with a single peak, the WKB formula is suitable for determining the dominant quasinormal modes The WKB method operates by matching asymptotic solutions, which adhere to the quasinormal boundary conditions (\ref{boundaryconditions}), with a Taylor expansion around the peak of the potential barrier. The first-order WKB formula, known as the eikonal approximation, becomes exact in the limit $\ell \to \infty$. Subsequently, the general WKB expression for frequencies can be expressed as an expansion around the eikonal limit, as follows \cite{Konoplya:2019hlu}:
\begin{eqnarray}\label{WKBformula-spherical}
\omega^2&=&V_0+A_2(\K^2)+A_4(\K^2)+A_6(\K^2)+\ldots\\\nonumber&-&\imo \K\sqrt{-2V_2}\left(1+A_3(\K^2)+A_5(\K^2)+A_7(\K^2)\ldots\right),
\end{eqnarray}
where the matching conditions for quasinormal modes imply that
\begin{equation}
\K=n+\frac{1}{2}, \quad n=0,1,2,\ldots,
\end{equation}
with $n$ representing the overtone number, $V_0$ denoting the maximum value of the effective potential, $V_2$ indicating the value of the second derivative of the potential at this maximum with respect to the tortoise coordinate, and $A_i$ for $i=2, 3, 4, \ldots$ signifying the $i$th WKB order correction term beyond the eikonal approximation, which is dependent on $\K$ and derivatives of the potential at its maximum up to the order $2i$. The explicit form of $A_i$ can be found in \cite{Iyer:1986np} for the second and third WKB orders, in \cite{Konoplya:2003ii,Konoplya:2004ip} for the 4th to 6th orders, and in \cite{Matyjasek:2017psv} for the 7th to 13th orders. The aforementioned WKB approach for determining quasinormal modes has been extensively employed at various orders in numerous studies (see, for instance, \cite{Abdalla:2005hu,Zinhailo:2019rwd,Konoplya:2006ar,Kokkotas:2010zd,Gonzalez:2022ote,Xia:2023zlf,Fernando:2014gda,Fernando:2012yw}).

\subsection{Time-domain integration}

Another way to find quasinormal modes of black hole is base on the time-domain integration, for which we employed the Gundlach-Price-Pullin discretization scheme \cite{Gundlach:1993tp}, expressed as:
\begin{eqnarray}
\Psi\left(N\right)&=&\Psi\left(W\right)+\Psi\left(E\right)-\Psi\left(S\right)\nonumber\\
&&- \Delta^2V\left(S\right)\frac{\Psi\left(W\right)+\Psi\left(E\right)}{4}+{\cal O}\left(\Delta^4\right),\label{Discretization}
\end{eqnarray}
where the integration scheme is defined by $N\equiv\left(u+\Delta,v+\Delta\right)$, $W\equiv\left(u+\Delta,v\right)$, $E\equiv\left(u,v+\Delta\right)$, and $S\equiv\left(u,v\right)$. This method has been extensively utilized in numerous studies \cite{Konoplya:2014lha,Konoplya:2005et,Abdalla:2012si,Varghese:2011ku,Qian:2022kaq}, demonstrating its accuracy.

\subsection{Prony method}

\begin{figure}
\resizebox{\linewidth}{!}{\includegraphics{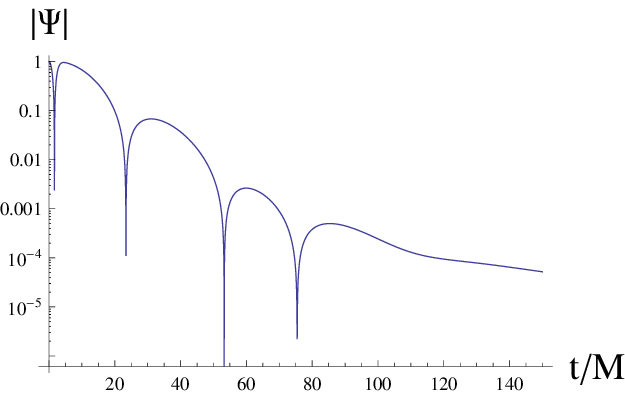}}
\caption{Time-domain profile for the scalar perturbations ($\ell=0$)  hole $\gamma = 0.5$, $M =1$.}\label{fig:timedomain}
\end{figure}

\begin{figure}
\resizebox{\linewidth}{!}{\includegraphics{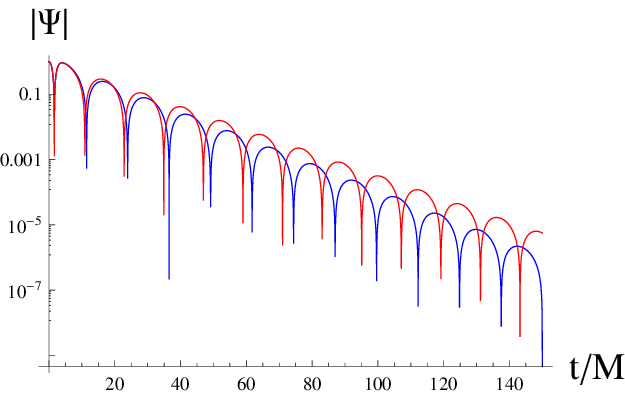}}
\caption{Time-domain profile for the electromgnetic perturbations ($\ell=1$) $\gamma =0.1 $ (blue) and $\gamma =1.6$ (red); $M =1$..}\label{fig:timedomain2}
\end{figure}

To extract the frequency values from the time-domain profile, we employ the Prony method, which entails fitting the profile data with a summation of damped exponents:
\begin{equation}\label{damping-exponents}
\Psi(t)\simeq\sum_{i=1}^pC_ie^{-i\omega_i t}.
\end{equation}
We assume that the quasinormal ringing phase initiates at some $t_0=0$ and concludes at $t=Nh$, where $N\geq2p-1$. Subsequently, equation (\ref{damping-exponents}) holds true for every point of the profile:
\begin{equation}
x_n\equiv\Psi(nh)=\sum_{j=1}^pC_je^{-i\omega_j nh}=\sum_{j=1}^pC_jz_j^n.
\end{equation}
We then determine $z_i$ based on the known $x_n$ and compute the quasinormal frequencies $\omega_i$. Quasinormal modes are typically derived from time-domain profiles when the ring-down phase comprises a sufficient number of oscillations. Additionally, the duration of the ringdown period increases with the multipole number $\ell$.
Typical time-domain profiles are shown here in figs. \ref{fig:timedomain} \ref{fig:timedomain2}. When $\ell=0$ (for scalar perturbations), the ringdown stage consists only of a few oscillations and extraction of frequencies are usually done with a relative error of a fraction of one percent or more. For higher multipoles, the frequencies can be extracted with sufficiently high accuracy. The Prony method has been recently used by the author in \cite{Skvortsova:2023zmj,Skvortsova:2023zca} for extracting quasinormal frequencies of BTZ-like black holes with various asymptotics \cite{Konoplya:2020ibi}. It turned out to be efficient not only for finding  the dominant modes, but also a few higher overtones \cite{Dubinsky:2024gwo}.

\begin{table*}
\begin{tabular}{c c c c c}
\hline
\hline
$\gamma $ & Prony fit & WKB6 Padé &  rel. error $Re (\omega)$ & rel. error $Im (\omega)$  \\
\hline
$0$ & $0.110035-0.104957 i$ & $0.110792-0.104683 i$ & $0.688\%$ & $0.261\%$ \\
$0.25$ & $0.111042-0.103680 i$ & $0.111632-0.103458 i$ & $0.532\%$ & $0.214\%$ \\
$0.5$ & $0.112037-0.102150 i$ & $0.111854-0.102217 i$ & $0.163\%$ & $0.0650\%$ \\
$0.75$ & $0.112963-0.100280 i$ & $0.112489-0.100448 i$ & $0.419\%$ & $0.168\%$ \\
$1.$ & $0.113677-0.097928 i$ & $0.113862-0.098605 i$ & $0.163\%$ & $0.691\%$ \\
$1.25$ & $0.113754-0.094890 i$ & $0.115918-0.094532 i$ & $1.90\%$ & $0.378\%$ \\
$1.5$ & $0.111735-0.091443 i$ & $0.112619-0.091029 i$ & $0.791\%$ & $0.453\%$ \\
$1.67$ & $0.109630-0.091320 i$ & $0.110774-0.091191 i$ & $1.04\%$ & $0.141\%$ \\
\hline
\hline
\end{tabular}
\caption{Comparison of the quasinormal frequencies obtained by the time-domain integration and the 6th order WKB approach with Padé approximants for $s=\ell=0$ ($M=1$).}\label{check1}
\end{table*}

\begin{table*}
\begin{tabular}{c c c c c}
\hline
\hline
$\gamma $ & Prony fit & WKB6 Padé & rel. error $Re (\omega)$ & rel. error $Im (\omega)$  \\
\hline
$0$ & $0.292945-0.097662 i$ & $0.292930-0.097663 i$ & $0.00524\%$ & $0.\times 10^{\text{-4}}\%$\\
$0.25$ & $0.294480-0.096599 i$ & $0.294471-0.096606 i$ & $0.00294\%$ & $0.0071\%$\\
$0.5$ & $0.296050-0.095371 i$ & $0.296047-0.095377 i$ & $0.00098\%$ & $0.0060\%$\\
$0.75$ & $0.297634-0.093940 i$ & $0.297628-0.093943 i$ & $0.00203\%$ & $0.0032\%$\\
$1.$ & $0.299186-0.092258 i$ & $0.299179-0.092262 i$ & $0.00243\%$ & $0.0039\%$\\
$1.25$ & $0.300611-0.090283 i$ & $0.300602-0.090291 i$ & $0.00292\%$ & $0.0089\%$\\
$1.5$ & $0.301744-0.088051 i$ & $0.301740-0.088069 i$ & $0.00112\%$ & $0.0208\%$\\
$1.67$ & $0.302305-0.086497 i$ & $0.302297-0.086505 i$ & $0.00262\%$ & $0.0087\%$\\
\hline
\hline
\end{tabular}
\caption{Comparison of the quasinormal frequencies obtained by the time-domain integration and the 6th order WKB approach with Padé approximants for $s=0$, $\ell=1$ ($M=1$).}\label{check2}
\end{table*}

\begin{table*}
\begin{tabular}{c c c c c}
\hline
\hline
$\gamma $ & Prony fit & WKB6 Padé & rel. error $Re (\omega)$ & rel. error $Im (\omega)$ \\
\hline
$0$ & $0.183033-0.096909 i$ & $0.182643-0.096566 i$ & $0.213\%$ & $0.354\%$\\
$0.25$ & $0.184043-0.095682 i$ & $0.183803-0.095315 i$ & $0.131\%$ & $0.384\%$\\
$0.5$ & $0.185030-0.094255 i$ & $0.184618-0.093923 i$ & $0.222\%$ & $0.352\%$\\
$0.75$ & $0.185940-0.092576 i$ & $0.185439-0.092454 i$ & $0.269\%$ & $0.132\%$\\
$1.$ & $0.186669-0.090582 i$ & $0.186289-0.090419 i$ & $0.203\%$ & $0.180\%$\\
$1.25$ & $0.186985-0.088228 i$ & $0.186918-0.087872 i$ & $0.0354\%$ & $0.404\%$\\
$1.5$ & $0.186434-0.085778 i$ & $0.186250-0.085328 i$ & $0.0988\%$ & $0.525\%$\\
$1.67$ & $0.185702-0.084646 i$ & $0.185543-0.084189 i$ & $0.0861\%$ & $0.540\%$\\
\hline
\hline
\end{tabular}
\caption{Comparison of the quasinormal frequencies obtained by the time-domain integration and the 6th order WKB approach with Padé approximants for $s=1/2$, $\ell=1/2$ ($M=1$).}\label{check3}
\end{table*}

\begin{table*}
\begin{tabular}{c c c c c}
\hline
\hline
$\gamma $ & Prony fit & WKB6 Padé & rel. error $Re (\omega)$ & rel. error $Im (\omega)$ \\
\hline
$0$ & $0.248266-0.092499 i$ & $0.248255-0.092497 i$ & $0.00449\%$ & $0.0030\%$\\
$0.25$ & $0.250197-0.091543 i$ & $0.250192-0.091552 i$ & $0.00192\%$ & $0.0096\%$\\
$0.5$ & $0.252227-0.090394 i$ & $0.252242-0.090399 i$ & $0.00587\%$ & $0.0048\%$\\
$0.75$ & $0.254350-0.088992 i$ & $0.254360-0.088987 i$ & $0.00427\%$ & $0.0059\%$\\
$1.$ & $0.256528-0.087246 i$ & $0.256525-0.087245 i$ & $0.00111\%$ & $0.0018\%$\\
$1.25$ & $0.258658-0.085046 i$ & $0.258647-0.085064 i$ & $0.00420\%$ & $0.0207\%$\\
$1.5$ & $0.260494-0.082336 i$ & $0.260503-0.082358 i$ & $0.00355\%$ & $0.0268\%$\\
$1.67$ & $0.261399-0.080327 i$ & $0.261448-0.080334 i$ & $0.0187\%$ & $0.0084\%$\\
\hline
\hline
\end{tabular}
\caption{Comparison of the quasinormal frequencies obtained by the time-domain integration and the 6th order WKB approach with Padé approximants for $s=1$, $\ell=1$ ($M=1$).}\label{check4}
\end{table*}

\begin{table*}
\begin{tabular}{c c c c c}
\hline
\hline
$\gamma $ & Prony fit & WKB6 Padé & rel. error $Re (\omega)$ & rel. error $Im (\omega)$ \\
\hline
$0$ & $0.457599-0.095002 i$ & $0.457596-0.095005 i$ & $0.00063\%$ & $0.0033\%$\\
$0.25$ & $0.460133-0.094010 i$ & $0.460131-0.094013 i$ & $0.00039\%$ & $0.0031\%$\\
$0.5$ & $0.462789-0.092864 i$ & $0.462788-0.092865 i$ & $0.00007\%$ & $0.0008\%$\\
$0.75$ & $0.465570-0.091524 i$ & $0.465569-0.091524 i$ & $0.00041\%$ & $0.0006\%$\\
$1.$ & $0.468473-0.089941 i$ & $0.468470-0.089941 i$ & $0.00067\%$ & $0.0004\%$\\
$1.25$ & $0.471474-0.088046 i$ & $0.471471-0.088048 i$ & $0.00067\%$ & $0.0021\%$\\
$1.5$ & $0.474510-0.085769 i$ & $0.474508-0.085770 i$ & $0.00048\%$ & $0.0005\%$\\
$1.67$ & $0.476542-0.083978 i$ & $0.476542-0.083975 i$ & $9.5\times 10^{-6}\%$ & $0.0032\%$\\
\hline
\hline
\end{tabular}
\caption{Comparison of the quasinormal frequencies obtained by the time-domain integration and the 6th order WKB approach with Padé approximants for $s=1$, $\ell=2$ ($M=1$).}\label{check5}
\end{table*}

\begin{table*}
\begin{tabular}{c c c c c}
\hline
\hline
$\gamma $ & Prony fit & WKB6 Padé & rel. error $Re (\omega)$ & rel. error $Im (\omega)$ \\
\hline
$0$ & $0.380042-0.096388 i$ & $0.380041-0.096408 i$ & $0.00015\%$ & $0.0207\%$\\
$0.25$ & $0.381879-0.095360 i$ & $0.381879-0.095380 i$ & $0.00011\%$ & $0.0210\%$\\
$0.5$ & $0.383776-0.094184 i$ & $0.383778-0.094201 i$ & $0.00057\%$ & $0.0177\%$\\
$0.75$ & $0.385719-0.092827 i$ & $0.385728-0.092840 i$ & $0.00227\%$ & $0.0141\%$\\
$1.$ & $0.387684-0.091250 i$ & $0.387690-0.091260 i$ & $0.00160\%$ & $0.0110\%$\\
$1.25$ & $0.389617-0.089411 i$ & $0.389620-0.089425 i$ & $0.00085\%$ & $0.0150\%$\\
$1.5$ & $0.391426-0.087298 i$ & $0.391432-0.087313 i$ & $0.00154\%$ & $0.0171\%$\\
$1.67$ & $0.392539-0.085733 i$ & $0.392546-0.085744 i$ & $0.00184\%$ & $0.0133\%$\\
\hline
\hline
\end{tabular}
\caption{Comparison of the quasinormal frequencies obtained by the time-domain integration and the 6th order WKB approach with Padé approximants for $s=1/2$, $\ell=3/2$ ($M=1$).}\label{check6}
\end{table*}

\section{Quasinormal modes}\label{sec:QNM}

Quasinormal modes of a test scalar field for the above quantum corrected black hole have been recently analyzed in \cite{Gong:2023ghh}. The scalar field frequencies were computed there with the three methods, the WKB, pseudo-spectral (PS) and Asymptotic Iteration (AIM) methods. The large discrepancy was detected there between the first method and the last two. For example, for $\ell =n=0$ modes of scalar perturbations they found  $\omega = 0.11467730 - 0.07528910i$ (WKB), $0.11191203 - 0.09141316i$ (PS), $0.11191397 - 0.09141256i$ (AIM). However, our calculations with the help of the 6th order WKB method with Padé approximants $\tilde{m}=4$ presented in table I shows a different value from their WKB data which is much closer to their PS and AIM data: $\omega = 0.112619-0.091029 i$ and the time-domain integration gives even closer results $\omega = 0.111735-0.091443 i$. The big error in the WKB data obtained in \cite{Gong:2023ghh} is apparently connected with the averaging over various choices of  $\tilde{m}$, which we do not make, but instead choose the best $\tilde{m}$ in the Schwarzschild limit, the precise value of quasinormal modes for which is known. 

The quasinormal modes are calculated here with the time-domain integration, and 6th order WKB approach combined with the Padé approximants.
Examples of the time-domain profile for the $\ell =0$ scalar and $\ell=1$ electromagnetic perturbations are given in figs. \ref{fig:timedomain} and  \ref{fig:timedomain2} respectively. From tables \ref{check1}- \ref{check6} we can see that the 6th order WKB method with the Padé approximants is in a very good concordance with the time-domain integration. Taking the time-domain integration data as accurate for the lowest multipole numbers we can see that the relative error usually does not exceed a small fraction of one percent. An exception is the $\ell=0$ scalar perturbations for which an error may reach one percent.  One can see that the overall effect, that is, the deviation of the quasinormal frequency from its Schwrazschild limit, is much bigger than the expected relative error for the WKB method, while the in the case $\ell=0$ we rely upon the time-domain integration results based on the converging procedure, more than on the WKB data.

From tables \ref{check1}- \ref{check6} we see that when the quantum parameter is tuned on, $Re \omega$ affected much less than the damping rate. In particular, we see that the real oscillation frequency is slightly increased (by several percents), while the damping rate, proportional to $Im \omega$, is considerably decreased. The exception from this behavior is $\ell=0$ scalar perturbations for near extreme black holes for which $Re \omega$ increases insignificantly. However, it is the regime in which the relative error is of the order of this insignificant growth and our conclusion is that the main effect of the non-zero quantum parameter is in considerably diminishing the damping rate of the perturbation.

From Tables \ref{check1}-\ref{check6}, we observe that when the quantum parameter is activated, $Re(\omega)$ is significantly less affected compared to the damping rate. Specifically, we note a slight increase (by several percent) in the real oscillation frequency, while the damping rate, proportional to $Im(\omega)$, experiences a considerable decrease. The only deviation from this trend occurs with $\ell=0$ scalar perturbations near extreme black holes, where the decrease in $Re(\omega)$ is insignificant. However, in this regime, the relative error is of the same order as this insignificant decrease, leading us to conclude that the primary effect of the non-zero quantum parameter is a significant reduction in the damping rate of the perturbation.

In the (eikonal) limit of high multipole numbers $\ell$, quasinormal modes corresponds to the parameters of the unstable null geodesics, Lyapunov exponent (for $Im (\omega)$) and angular velocity (for $Re (\omega)$) \cite{Cardoso:2008bp}. 
Thus, one has 
\begin{equation}\label{QNM}
\omega_n=\Omega\ell-\imo(n+1/2)|\lambda|, \quad \ell \gg n,
\end{equation}
where, $\Omega$ is the angular velocity at the unstable circular null geodesics, and $\lambda$ is the Lyapunov exponent.
However, the correspondence may be broken in a number of cases \cite{Konoplya:2017wot,Khanna:2016yow,Konoplya:2022gjp,Bolokhov:2023dxq,Konoplya:2020bxa} and the check is necessary for every case. The eikonal limit of quasinormal frequencies was extensively investigated in a great number of works for various configurations \cite{Konoplya:2018ala,Paul:2023eep,Bolokhov:2022rqv,Konoplya:2005sy,Bolokhov:2023bwm,Zhidenko:2008fp,doi:10.1142/S0217751X24500246,Dolan:2010wr,Hod:2009td,Jusufi:2020dhz,Breton:2016mqh}.

An unstable null circular geodesic at a radius $r=r_c$ satisfies the condition $U'(r)\bigl|_{r_c}=0$, where $U$ is an effective potential for the particle motion. Then according to \cite{Cardoso:2008bp} we have
\begin{equation}\label{rccond}
2f_c-r_c f'_c=0,
\end{equation}
and therefore, the real oscillation frequency obeys formally the same equation as the first order WKB formula.
The angular velocity $\Omega=d\varphi/dt=\dot{\varphi}/\dot{t}$ can be obtained using the circular motion condition $\dot{r}=0$:
\begin{equation}\label{Omegac}
\Omega=\frac{\sqrt{f_c}}{r_c}.
\end{equation}

\begin{figure*}
\resizebox{\linewidth}{!}{\includegraphics{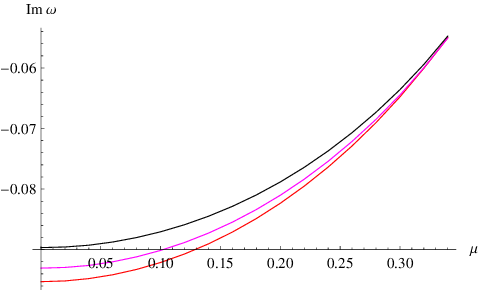}\includegraphics{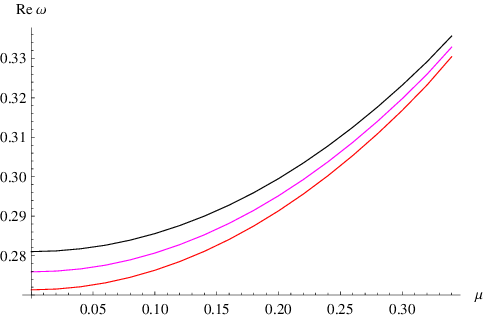}}
\caption{Quasinormal modes for the massive scalar field perturbations ($\ell=1$) $\gamma =0.01 $ (red),  $\gamma =0.5$ (violet), $\gamma =1$ (black); $M =1$..}\label{fig:Massive}
\end{figure*}

Using the expansion in terms of the inverse multipole moment we can find the position of the maximum of the effective potential in the eikonal limit,
\begin{equation}
r_{max} = \left(3 M-\frac{\gamma }{9 M}+O\left(\gamma
   ^2\right)\right)+\frac{O\left(\gamma ^2\right)}{\kappa
   }+O\left(\left(\frac{1}{\kappa }\right)^2\right),
\end{equation}
and the quasinormal frequencies in the eikonal limit,
\begin{widetext}
\begin{equation}
\omega = \kappa  \left(\frac{1}{3 \sqrt{3} M}+\frac{\gamma }{162
   \sqrt{3} M^3}+O\left(\gamma
   ^2\right)\right)+\left(-\frac{i K}{3 \sqrt{3} M}+\frac{i
   \gamma  K}{81 \sqrt{3} M^3}+O\left(\gamma
   ^2\right)\right)+O\left(\left(\frac{1}{\kappa
   }\right)^1\right).
\end{equation}
\end{widetext}
Here $\kappa =\ell +\frac{1}{2}$, $K= n+\frac{1}{2}$. When $\gamma =0$, the above expressions are reduced to the well-known expressions for the Schwarzschild quasinormal frequencies.
From the comparison with the parameters of the unstable null geodesics it can be easily seen that the correspondence holds in this case. 

Finally, we would like to discuss the spectrum of a massive term using the example of a scalar field. While massive fields are typically short ranged, there is at least a theoretical possibility of observing them in experiments involving very long gravitational waves, such as recent Pulsar  Timing  Array experiments \cite{NANOGrav:2023gor,NANOGrav:2023hvm,EuropeanPulsarTimingArray:2023egv}. Not only are quasinormal modes long-lived, but their asymptotic tails could also potentially be observed \cite{Konoplya:2023fmh}. Furthermore, there are additional motivations for studying the spectrum of massive fields. For instance, massless fields in brane-world scenarios and higher-dimensional gravity may manifest in our \$(3+1)\$- brane as effectively massive \cite{Seahra:2004fg,Ishihara:2008re}, while a magnetic field in the vicinity of a black hole can also induce an effective massive term \cite{Konoplya:2008hj}. 

The real and imaginary parts of the quasinormal frequencies for a massive scalar field at various values of the coupling $\gamma$ calculated by the 6th order WKB method with Padé approximants  are shown in fig. \ref{fig:Massive}. There one can see that when the mass is turned on, the real oscillation frequency is increased, while the damping rate is decreasing. Morevoer, the damping rates for various $\gamma$ approach each other and tend to coincide with the Schwarzschild one. At the same time, it is well-known that for the Schwarzschild black hole a massive scalar field, as well as fields of other spin,  contain the so-called {\it quasi-resonances} \cite{Ohashi:2004wr}, which have been extensively studied in \cite{Bolokhov:2023bwm,Konoplya:2019hml,Bolokhov:2023ruj,Konoplya:2017tvu,Zinhailo:2018ska}. These modes have arbitrarily small damping rate when $\mu$ approaches a certain critical value.  Nevertheless, there are exceptions when quasi-resonances do not exist \cite{Zinhailo:2024jzt}. From fig. \ref{fig:Massive} we conclude that they exist and coincide with the Schwarzschild ones when the quantum correction is turned on. We cannot extend calculations of the massive case to larger $\mu$, because for sufficinetly large mass of the field, the effective potential does not have a maximum anymore and the whole WKB approach is inapplicable. As noticed in  \cite{Konoplya:2019hlu}, the higher WKB method with Padé approximants still provides good estimates in the regime of relatively small masses $\mu$.

\vspace{3mm}
\section{Conclusions}\label{sec:conclusions}

In this paper, we have computed the dominant quasinormal modes for scalar, electromagnetic, and Dirac (neutrino) perturbations of a quantum-corrected black hole developed in \cite{Lewandowski:2022zce}. Comparison between time-domain integration and the 6th-order WKB method with Padé approximants reveals excellent agreement. We expand upon previous work \cite{Gong:2023ghh}, which focused solely on scalar perturbations of this black hole, by including neutrino and electromagnetic fields. Our findings demonstrate that employing appropriate Padé approximants can significantly enhance the consistency between the WKB method and more precise techniques. 
In addition, we have derived the analytic formula for the quasinormal limit in the eikonal limit and showed that it respects the null geodesics/eikonal quasinormal modes correspondence. 
For a massive scalar field we have demonstrated the existence of arbitrarily long lived quasinormal modes, quasi-resonances. Future research may extend these calculations to explore the grey-body factors of the quantum-corrected black hole further.

\vspace{2mm}
\begin{acknowledgments}
I would like to acknowledge R. A. Konoplya for useful discussions and patience. This work was supported by RUDN University research project FSSF-2023-0003.
\end{acknowledgments}

\bibliographystyle{unsrt}
\bibliography{bibliography}
\end{document}